\documentclass[12pt]{article}
\usepackage{amsmath}
\usepackage[dvips]{graphicx}
\usepackage{epsfig}
\usepackage{amsmath}
\usepackage{amssymb}
\usepackage{float}
\usepackage{graphics,epstopdf}
\usepackage{amsfonts}
\usepackage{amsthm}
\usepackage{longtable}
\usepackage{tabularx}
\usepackage[usenames]{color}
\definecolor{AV}{rgb}{0.65,0.0,0}
\definecolor{GC}{rgb}{0,0.0,0.65}
\definecolor{WS}{rgb}{0,0.65,0}

\usepackage[
      colorlinks=true,
      linkcolor=blue,
      urlcolor=blue,
      filecolor=blue,
      citecolor=blue,
      pdfstartview=FitV,
      pdftitle={},
      pdfauthor={},
      pdfsubject={},
      pdfkeywords={},
      pdfpagemode=None,
      bookmarksopen=true
]{hyperref}

\newcommand{\bm}{\begin{multiline}}
\newcommand{\beq}{\begin{equation}}
\newcommand{\eeq}{\end{equation}}
\newcommand{\beqs}{\begin{eqnarray}}
\newcommand{\eeqs}{\end{eqnarray}}

\begin{document}

\thispagestyle{empty}

\hfill{}

\hfill{}

\hfill{}

\vspace{32pt}

\begin{center}

\textbf{\Large Statistical distributions of mean motion resonances and near-resonances in multiplanetary systems}

\vspace{48pt}

\textbf{Marian C. Ghilea}\footnote{E-mail: \texttt{mghilea@gmail.com}}

\vspace*{0.2cm}

\end{center}

\vspace{30pt}

\begin{abstract}
 The orbits of the confirmed exoplanets from all multiple systems known to date are investigated. Observational data from 1890 objects, of which 1176 are found in multiplanetary systems, are compiled and analyzed. Mean motion resonances and near-resonances up to the outer/inner orbital period ratio's value of 5 and the denominator 4 are tested for all adjacent exoplanet orbits. Each host star's ``snow line'' is calculated  using a simple algorithm. The planets are reclassified into categories as a function of the semimajor axis size relative to the snow line location and the semimajor axis vs mass distribution. The fraction of planets in/near resonance is then plotted as a function of  both resonance number and resonance order for all the exoplanet population and, separately, for each planet type. In the resonance number plot it appears that the 2/1 and 3/2 resonances and near-resonances are dominant overall and for the giant planets, but the observed distribution profile changes significantly with each planet category, with terrestrial planets, neptunes and mini-neptunes showing the largest variation. Resonances/near resonances around the value 5/3 were dominant for mini neptunes and terrestrial planets. In the order-based resonance/near-resonance plot, the observed distribution appears to follow an exponential decay for the general population and its profile appears to be influenced by the planet type. Approximate methods to estimate resonance/near resonance distributions are also attempted for the systems with unknown planet mass or with unknown star and/or planet mass and compared with the distribution of the planets with all the parameters known. A separate study of the resonance/near resonance fraction distribution as a function of mass is also attempted, but the low statistical data at very high planetary masses prevent the finding of an accurate equation to describe such a dependency.
\end{abstract}
\vspace{32pt}

\setcounter{footnote}{0}

\newpage

\section{Introduction}

Since the discovery of the first exoplanet in 1995 around the star 51 Pegasi by Mayor and Queloz \cite{Mayor:1995eh}, a plethora of planetary bodies orbiting other stars was found and this category of celestial objects has been increasing in size at an almost exponential pace. The number of such objects discovered by both terrestrial and space-based observation is rapidly approaching 2000, with over 5000 candidates in existence. Radial velocity \cite{Mayor:1995eh} and astrometry \cite{1996AAS...188.4011G} remain the main indirect detection methods used to identify low magnitude perturbations in the motion of the host star while transit detections \cite{charbonneau2000detection}, microlensing \cite{bond2004ogle} or direct imaging  \cite{lagrange2009probable} seek to reveal new planets through direct photometric, gravitational or optical effects generated by their presence. The fraction of stars where planetary systems are present is in general estimated from both theoretical observations and observational data to be around 30-40\% \cite{{tutukov2012formation}, {borucki2011characteristics},  {eisner2008proplyds}, {roberge2008debris}}.\\
A significant number of the newly discovered exoplanets (more than 60\%) are part of multiple planetary systems, and, as the detection sensitivity continues to increase, they will most likely become the norm. Based on their mass, density, and semimajor axis, they are typically classified into several notable categories:\\
a)  hot jupiters  - giant planets with masses comparable to Jupiter's size (probably with a small rocky core surrounded by metallic hydrogen and a thick atmosphere), conventionally taken as 100 times the mass of Earth ($M_E$) or more, orbiting very close to the host star, at distances where their surface temperatures are of the order of hundreds to thousands of Celsius degrees \cite{Mayor:1995eh};\\
b) jupiters - planets similar to Jupiter in appearance and surface temperature, normally found near or beyond the host star's ``snow line" \cite{akeson2013nasa};\\
c) neptunes - objects that resemble Neptune, with masses of 10-100 $M_E$, most often made of a small rock core surrounded by a larger ice mantle (water, ammonia or methane) and a thick atmosphere; normally found beyond the star's ``snow line" \cite{akeson2013nasa} ;\\
d) hot neptunes - giant planets similar to Neptune, with masses of 10-100 $M_E$, orbiting the host star at a distance that rises their surface temperature to several hundreds Celsius degrees or more \cite{Santos:2004yu};\\
e) terrestrial-type planets - rocky objects surrounded by a thin atmosphere, with masses up to 10 $M_E$, a significant fraction of them consisting of super-earths \cite{Valencia:2006fx}; the terrestrial planets positioned into the habitable zone are of particular importance, as possible hosts for life;\\
f) mini-neptunes - gas dwarfs with a mass comparable with that of terrestrial planets (i.e. less than 10 $M_E$), but with a low density; if their atmosphere also has a low density this might be due to a significant water layer, several thousands km deep and they can be classified as ocean planets \cite {Barnes:2009gt};\\
g) brown dwarfs - objects at the boundary between stars and planets, most often considered by convention as such if their masses are larger than 13 Jupiter masses (or about 4000 $M_E$), the lowest value at which the deuterium fusion is possible \cite {basri1996lithium}.\\ 
The above-used classification is not exhaustive, with some possible cases in-between, for example terrestrial-type planets (super-Earths) with masses larger than 10 ($M_E$). However, statistically, such a classification appears to be accurate in most situations.\\ 
The causes that determine the distribution of mass and momentum of the orbiting bodies in multiplanetary systems are still incompletely understood and depend on many parameters, including the host star type and rotation speed, the initial mass of the protoplanetary disk and its distribution, external gravitational perturbations, chemical composition of the initial nebula mass and planet migration. For this reason, the subsequent stable or quasi-stable configuration of a planetary system might give insights to these initial conditions and fuel a better understanding of the processes that generate the planet formation.\\
After a planetary system has formed, the combined work of gravitational perturbations of all the orbiting bodies push it towards certain structure types. If these perturbations are satisfactorily quantified, the composition and dynamics of the system can be at least partly explained. The analysis of the apparition and propagation of perturbations in multiplanetary systems requires solving the classical Newtonian N-body problem. With suitable approximations it is possible to find analytical solutions to particular configurations and these solutions can be used for systems with two or more planets. This can be done by considering the effects of the purely secular terms in the disturbing function for a system of n masses orbiting a central body - also known as the secular perturbation theory.\\
 For certain cases, where the ratio of two planets' orbital periods (or characteristic frequencies) is close to a ratio of integer numbers, they are said to be in or near resonance. For the resonance case, the periodic oscillation motion resembles that of a pendulum, inducing libration while the relative planet positions swing back and forth during an integer number of complete orbits. During the motion out of the resonance range, the libration transforms into circulation with the two modes separated by an apsidal separatrix. The boundary between libration (either aligned or anti-aligned) and circulation forms the simplest such type of separation boundary. Apparently, many of the known orbital systems lie close to apsidal separatrixes \cite{{Ford:2005ta}, {Barnes:2006aa}, {Barnes:2006cn}}. \\
 If significant and relevant statistical data based on observations displaying the fraction of each planet type that is near a mean motion resonance can be gathered, it can be potentially tested either analytically or numerically to predict possible configurations in incompletely known systems. It can also generate a more detailed model of the structure and formation of multiplanetary systems. Such a statistical study of the mean motion resonance and near-resonance occurrences in multiplanetary systems, using the available data from the confirmed exoplanets, makes the purpose of this article. \\

\section{Planetary mean-motion resonances}

The planetary resonances are thought to have an important role in defining the ultimate structure of a planetary system. They appear when the characteristic frequencies of two or more  bodies orbiting a central, much larger mass, are close to an exact commensurability \cite {{murray1999solar}, {armitage2010astrophysics}}. When such a situation occurs, the gravitational forces exerted by planetary motions add up over time in a coherent manner. \\
The simplest type of resonance is the mean-motion resonance and takes place when the orbital periods $P_1$ and $P_2$ of two planerts satisfy a relation of the type:
\begin{equation}       	
{P_2 \over P_1} \approx {j \over i},
\end{equation}
with j and i positive integers. Any mean-motion resonance has a well-defined width inside which the two orbiting bodies can have a libration movement around an equilibrium point.\\
 Planetary resonances or near-resonances are also thought to explain some regularities in the orbital distribution of the planets in multiplanetary systems, including the so-called ``Titius-Bode'' rule \cite{{roy1954occurrence}, {patterson1987resonance}}. Murray and Dermott demonstrated in 1999 that most of the random systems of bodies orbiting a central mass after achieving a relatively stable configuration would obey a distribution type that fits a Titius-Bode pattern \cite{murray1999solar}. \\
Winter and Murray \cite{{winter1997resonance}, {winter1997resonance2}} have classified the planetary mean-motion resonances into internal and external  as a function of the dominant mass position relative to the reference body in resonance. While the internal resonances (with the dominant mass on an external orbit) have analytical solutions for restricted cases of planar, circular three-body problem, the external resonances are rather described numerically or using the Hamiltonian approach. However, using these other methods makes the effect of the resonance on the orbital elements difficult to visualize and quantify. \\
	Without insisting on the basic theory, described in detail in many previous works  (such as, but not limited to \cite{{murray1999solar}, {winter1997resonance},{winter1997resonance2}}), it is worth mentioning that the variation of the orbital elements of two planets moving around the same star for a given potential can be described by the Lagrange's equations, where the variations of the orbital elements determine the system's motion over time. From here, the orbital evolution of a test particle can be studied by direct integration of the equations of motion \cite{{winter1997resonance}, {brouwer1961methods}, {dermott1983nature}}. \\
Using the pendulum model approach, an analytical solution can be derived for an interior resonance of the form:
\begin{equation}   
{(p+q)n'}  \approx {pn},
\end{equation}
with $p$, $q$ integers, $n$ and $n'$ - the mean motions of the two orbiting bodies. In this case, the libration width of the test particle can be calculated directly \cite{dermott1983nature}. For the first order resonances, it can be expressed using the semimajor axis relative variation in the equation:
\begin{equation}       	
{\delta a_{max} \over a}= \pm \Bigg( {16 \over 3} {{ |C_r| e}\over n}\Bigg)^{1/2} \Bigg( 1+ {1 \over {27 j_2^2 e^3}}  { |C_r| \over n}\Bigg)^{1/2} -{2 \over {9 j_2 e}  } { |C_r| \over n},
\end{equation}
or for the variation of the mean motion:
\begin{equation}    
{\delta n_{max} \over n} = \pm \Bigg({{12 |C_r| e} \over n}\Bigg)^{1/2} \Bigg( 1+ {1 \over {27 j_2^2 e^3}}  { |C_r| \over n}\Bigg)^{1/2} + { |C_r| \over{3 j_2 n^2}}.
\end{equation}
Here $a$ is the semimajor axis, $e$ is the orbit eccentricity, $n$ the mean motion, $j_2=-p$.\\
For the higher-order resonances, the expression changes into the simpler form \cite{murray1999solar}:
\begin{equation}    
{\delta n_{max} \over n} = \pm \Bigg({{12 |C_r| e^{q}} \over n}\Bigg)^{1/2},
\end{equation}
with $q$ the order of resonance.
$C_r$ is described by:
\begin{equation}   
{C_r} = {(-1)^{q}{ {m'} \over {M} }n \alpha f(\alpha)},
\end{equation}
where $m'$ is the mass of the perturber and $M$ the mass of the central object (the host star). The function $f(\alpha)$ is the disturbing function, with $\alpha$ the semimajor axis ratio for the two bodies in resonance. It can be expanded as a combination of Laplace coefficients, depending on the resonance order. \\
The expressions for $f(\alpha)$  are in fact already calculated in detail in ``Solar System Dynamics'' by Murray and Dermott  \cite {murray1999solar}, where they are described up to the 4th order resonances. Once the expression for the disturbing function is known analytically for the resonance type of interest, its numerical value can be easily and accurately found through well known computational algorithms that have been in use for over half a century \cite{izsak1963laplace}.\\
If the semimajor axis or mean motion relative difference is within the libration width, one can estimate that the two planets are in mean-motion resonance. Using an analytical expression for these values, as in Eq. (3) or (4), simplifies and speeds up considerably a computational analysis algorithm for a large database of planets but can generate errors, especially when the internal orbiting body has a larger mass and the resonance can be better described as external. However, this difficulty can be largely avoided if the above method of calculation is used to estimate that a system is either in a resonant or a near-resonant state. The method can be then tested against the many known cases of the exoplanet resonances that had been thoroughly verified in the existing literature and the libration width for the systems that are considered near-resonant can be increased until the analytical calculations agree with other, more complex, methods of calculation and the available observational data.

\section{The snow line}

As already shown in the introductory section, the classification of planets depends strongly of their orbital parameters relative to the ``snow line'', seen as the boundary where the water particles orbiting the host star switch from the state of vapors to ice. Typically, the temperature where the vapors-ice switch occurs is considered to be in the range of 145-170 K \cite{podolak2004note}.  Even though some highly eccentric orbits that cross this boundary may be encountered sometimes, the exoplanets are classified as inside or outside the snow line using their semi-major axis as a criterion for comparison. A way to find the value of this boundary in a fast and simple way (necessary to analyze databases of $10^3 - 10^4$ stars) but also sufficiently accurately to be used in a general planet classification is presented in this section. \\
The snow line is calculated approximately for each stellar system using the results from the work of Kennedy and Kenyon \cite{kennedy2008planet} (a detailed explanation of the method can be found in their article). If a thin accretion disk is considered, its mid temperature during the pre-main sequence evolution can be written as a sum of its mid-plane temperature resulting from viscous forces within this disk and the contribution from the star irradiation:
\begin{equation}       	
T_{mid}^4=T_{mid-accr}^4+T_{irr}^4,
\end{equation}
\noindent
Hubeny \cite {hubeny1990vertical} expresses the accretion temperature of the mid-plane as:
\begin{equation}       	
T_{mid-accr}^4 \sim {3\tau T_{eff-accr}^4 \over 8},
\end{equation}
\noindent
where $\tau=k \sigma_g/2$, with $k$ being opacity of the disk and a function of temperature \cite{bell427c} and $\sigma_g$ the surface gas density (about 100 times greater than the density of the dust \cite {natta2000protostars}). The parameter $T_{eff-accr}$ is expressed, in its turn by \cite{LyndenBell:1974kk}:
\begin{equation}       	
T_{eff-accr}^4 \sim {3 \over 8\pi}{{GM_{S}\dot{M}} \over {\sigma_{sb}a^3}}\Bigg(1-\Bigg({R_S \over a}\Bigg)^{1 \over 2}\Bigg),
\end{equation}
\noindent
where $\sigma_{sb}$ is the Stefan-Boltzmann constant, $M_S$ the star mass, $\dot{M}$ the accretion rate, $a$ - the snow line (expressed in AU) and $R_S$ the star radius. The value of  $\dot{M}$ is normally set to $\sim 10^{-8} M_S /yr$ \cite{{hartmann1998accretion}, {Kenyon:1987js}, {Chiang:1997it}}. \\
The irradiation contribution from the disk temperature is described by the relation:
\begin{equation}       	
T_{irr}^4 = T_S^4 \Bigg({\alpha \over 2} \Bigg ) \Bigg({R_S \over a} \Bigg )^3 ,
\end{equation}
\noindent 
with $\alpha \approx 0.005/a + 0.05 a^{2/7}$ for a flared disk in vertical hydrostatic equilibrium \cite{adams1986infrared}. For an optical thick disk, its temperature can be approximated with $T_{irr}$  \cite{Chiang:1997it}. Once $T_{mid}$ is set to 170 K as the preferred temperature for vapors-ice switch \cite{kennedy2008planet}, the value of $a$ can be calculated from Eq. (10).\\
A different approach to the calculation of the snow line is used by Ida \& Linn \cite{Ida:2005dg}. They have found the value of the snow line using the main sequence star luminosity ($L_s \propto M_S^4$) and an optically thin disk ($ T_{disk}^4 \propto L_S a^{-2}$) and obtaining a dependency of the type:
\begin{equation}       	
a_{snow} = 2.7 { M_S \over M_{Sun}} .
\end{equation}
Most literature estimates put the snow line for the Solar System at about 2.7 AU during the time of planet formation \cite {{armitage2010astrophysics}, {morbidelli2000source}}. The most compelling evidence for this value comes from the water-rich carbonaceous chondrite meteorites with the reflectance spectra matching the objects found in the outer asteroid belt beyond 2.5 AU, while meteorites coming from the inner belt (at around 2 AU) have negligible amounts of water \cite{morbidelli2000source}. \\
\begin{figure}[htb!]
  \centering
  \includegraphics [width=0.9\textwidth]{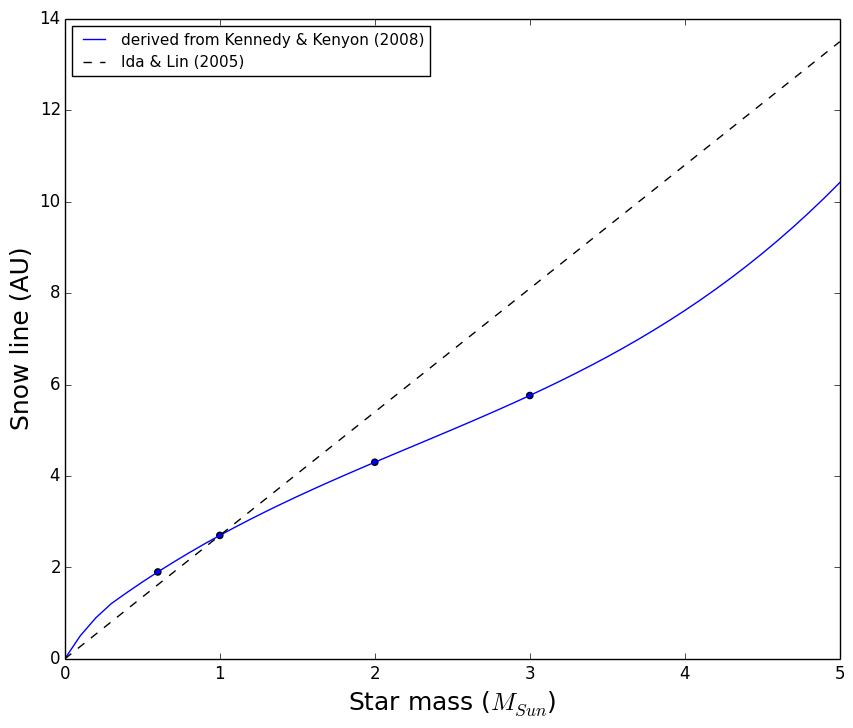}\\
  \caption{\emph{The snow line location as a function of the host star's mass. The curved line calculation is derived from the work of Kennedy and Kenyon. The straight line comes from the approach used by Ida and Lin. The four dots are points calculated by Kennedy and Kenyon. }}\label{Snow.png}
\end{figure}
\noindent
The snow line has a constant value for any host star if  Eq. (11) is applied, but it will change over time when it is calculated using the value of $T_{mid}$ from Eq. (7). Most of the literature sources use values of $\sim10^5$ years, which is approximately the time needed for the formation of the giant planets \cite{{podolak2004note}, {kennedy2008planet}}. If Eq. (7) is used for a star of one solar mass, the 2.7 AU value of the snow line is reached after $10^{5.6}$ years \cite{kennedy2008planet}. Calculating accurately the snow line is still a laborious process, necessitating the pre-main-sequence tracks for every involved star mass \cite{kennedy2008planet}. However, for the temporal value of $10^{5.6}$ years a simple polynomial function can be used to fit a snow line (measured in AU) with the data provided by Kennedy and Kenyon and to derive an approximate mass dependency for any stellar system described by the equation below:
\begin{equation}
   a_{snow} = \begin{cases}
               0.09 M_S^{3} + 0.609 M_S^{2} + 2.799 M_S +0.421 $ if $ M_S \ge 0.3,\\
               4.762 M_S^{3} - 7.093 M_S^{2}+ 5.71 M_S $ if $ M_S < 0.3,
           \end{cases}
\end{equation}
where $M_S$ is expressed in Sun masses ($M_{Sun}$).\\
\noindent
When $M_S > 0.3$, a function derived directly from  Kennedy \& Kenyon is applied \cite{kennedy2008planet}, while for $M_S < 0.3$, a different function was empirically calculated, to fit the reference data at $M_S =  0.3$ and to reach the origin as the star mass decreases to 0. \\
 Eq. (12) is used in the next section to rapidly calculate the snow line for over a thousand stars surrounded by planetary bodies with masses varying from 0.02 to 5 solar masses. The plot of the resulting calculation using this method and also the line from the simple equation used by Ida and Linn \cite{Ida:2005dg}, shown for comparison, are displayed in Figure 1.

\section{An adjusted classification of the exoplanets, based on the new calculation of the snow line and observational data}

The importance of the snow line location in planet formation cannot be neglected and it shall be used further as a calibrator to divide the known exoplanets in categories.  From the values of the snow line calculated with the method described in the previous section, a logarithmic density plot of the planets was derived as a function of planet mass and the semimajor axis/snow line ratio (Figure 2), for all the discovered exoplanets (Figure 2a) and, separately, only for the planets found in multiple systems (Figure 2b). From Figure 2a it can be seen that the planets tend to group in three main categories: \\
a) objects with large masses (over 100 $M_E$) and orbiting very close to the host star (semimajor axis less than 10\% of the snow line value) - the hot jupiters;\\
b) objects with slighly higher masses than the first category (most of them over 200 $M_E$) orbiting at about 10\%-200\% of  the host star's snow line - the classical giant planets or jupiters;\\
c) bodies with masses between 0.9 $-$ 20$M_E$, orbiting the host star inside the snow line (the terrestrial and some mini-neptune and neptune-type planets). This last category has a lower density plot, that comes mainly from the multiplanetary systems.\\
The neptune or hot neptune-type planets are located in between these categories and occur less often than the main three categories.
If only the objects from multiple systems are plotted (Figure 2b), the hot jupiters and some of the cooler giant planets disappear from the overall distribution graph. This fact proves that the hot-jupiter type objects have most likely migrated from the outside boundary of the snow line and absorbed most of the planet-making material from the initial nebula, preventing other sufficiently large planets from forming \cite{tsiganis2005origin}. The density plot for multiplanetary systems is left with two very loosely defined areas of terrestrial planets and cool giants with the intermediate-mass ice giants from the Neptune category almost filling the gap. The  planet masses appear to have a slight dependence of the snow line, tending to increase with the semimajor axis measured in snow line units. However, this should be regarded with caution, as the sensitivity of the observational methods in use is much higher for massive planets at large distances from the host star than for intermediate and low-mass objects. \\
\noindent 
Using the density plots from Figure 2, the planet classification presented in the introductory section was adjusted as following:\\
i) brown dwarfs - objects with masses over 4000 $M_E$ (or 13 Jupiter masses);\\
ii) hot jupiters - objects with masses between 100 and 4000 $M_E$ and with a semimajor axis less than 10\% of the snow line value;\\
iii) jupiters - same mass range as the hot jupiters, but with a semimajor axis larger than 10\% of the snow line;\\
iv) hot neptunes - objects within a mass range of 10 $-$ 100 $M_E$ and the semimajor axis less than 10\% of the snow line;\\
v) neptunes - objects within a mass range of 10 $-$ 100 $M_E$ and the semimajor axis larger than 10\% of the snow line;\\
\begin{figure}[H]
  \centering
  \includegraphics [width=0.70\textwidth]{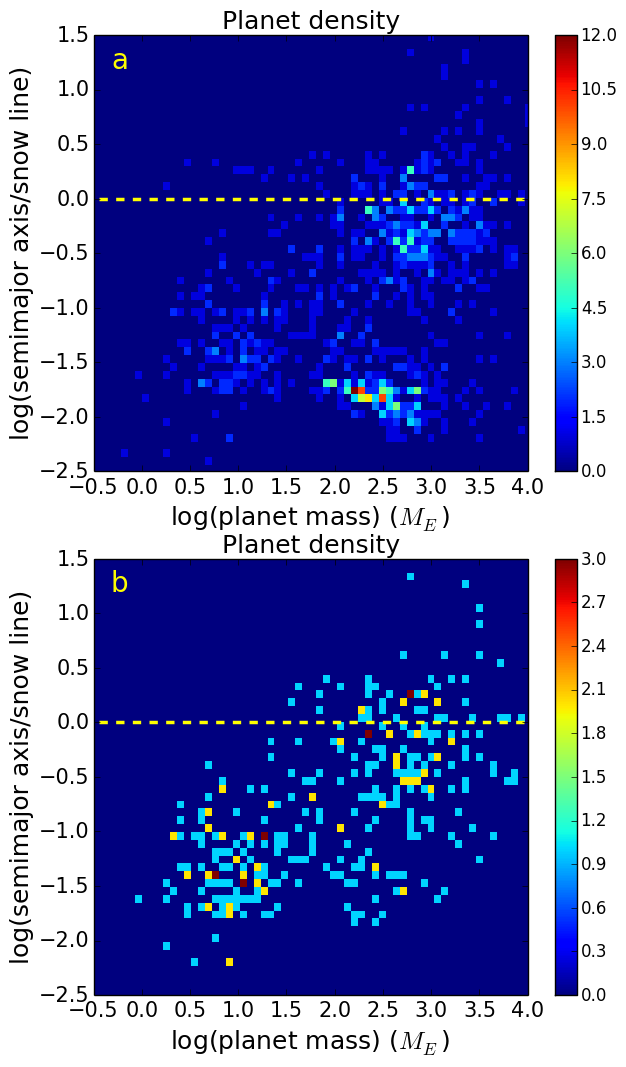}\\
  \caption{\emph{Density plot of planets as a function of mass (measured in $M_E$) and normalized snow line for all discovered exoplanets (a) and the objects present in multiplanetary systems (b). The snow line is represented by the yellow dashed line.}}\label{Mass-Snow.png}
\end{figure}
\noindent
vi) mini neptunes - objects with a mass less than 10 $M_E$ but with a low density (less than 3 g/$cm^3$) - some of these planets could also be ocean planets;\\
vii) terrestrial planets - objects with masses up to 10 $M_E$ and the semimajor axis less than the snow line;\\
viii) ice-terrestrial planets (or ice planets) - bodies with masses up to 10 $M_E$, but orbiting beyond the snow line and, in consequence expected to be composed mainly of frozen water, methane or ammonia; it is interesting to observe that so far no such planets have yet been discovered in multiple systems (most likely due to their high orbital period and the current limits of the detection sensitivity).\\
It is well known that during the process of planet formation the giant planets begin to take shape outside the snow line boundary and almost all of them experience an inward migration process \cite{{armitage2010astrophysics}, {tsiganis2005origin}}. As it can also be seen in Figure 2a, the vast majority of the known giant planets appear to have migrated inwards inside the snow line during their formation. The added category of ``ice planets'' has a low number of samples due to the small mas and large distance from the host star, which makes them difficult to detect with the current instruments. Being situated beyond the snow line, it is most likely that they are built from materials with low evaporation points that had previously migrated from the inside of the stellar system.

\section{Orbital period ratio resonance values and the frequency of their occurrences}

With the relatively high number of already discovered exoplanets, a general analysis of their orbital properties can be performed to determine statistically significant correlations between the mean motion resonance number or order for planets with adjacent orbits and the planet type. In order to achieve this, software packages such as  Python \cite{van2009python} and Matplotlib \cite{Hunter:2007}  were used for data analysis and respectively data plotting because of their high efficiency in processing of large amounts of information. Data sets from http://exoplanet.eu  (L'Observatoire de Paris) and http://exoplanetarchive.ipac.caltech.edu (NASA Exoplanet Archive) were downloaded and correlated into a single, combined catalog. While the first source was used as the main reference, data from the second website, related to planet density, star radius, stellar rotation and stellar activity was extracted and appended to it.\\
 The exoplanet detection method was neglected in this study. For example, it is well known that while the radial velocity technique measures the $m \sin i$ value, with $m$ the planet mass and $i$ the orbital angle, it would only provide the distribution of $m\sin i$, but because this distribution for a random set of planetary systems is very close to the $m$ distribution \cite{jorissen2001distribution}, the mass parameters were used as displayed in the source  databases. \\
From a total of 1890 exoplanets discovered to date, 1166 planets are part of 479 multiple planetary systems. All these objects from multiple systems were analyzed and automatically classified in the eight categories described in the previous section. The planets where both the star mass and planet mass were known were classified as ``defined''. The cases where the star mass was known, but not the planet mass, were classified as ``undefined'' while those where the star mass was unknown as ``unknown'' (as shown in Table 1).\\
\begin{table}
  
\begin{center}
    \begin{tabular}{ p{3.0cm} p{2.0cm} p{3.1cm} p{2.0cm} p{2.6cm}}

    \hline
    
\textbf{planet type} &\textbf{total} &	\textbf{in multiplanet \newline systems} & \textbf{near/ \newline on res.} & \textbf{fraction near/on res.} \\ \hline
all types &	1890 &	1166 &	330 &	0.28\\
defined &	1042 &	385 &	156 &	0.41\\
brown dwarf &	58 &	12 &	5 &	0.42\\
hot jupiter &	295 &	48 &	29 &	0.6\\
jupiter &	408 &	137 &	58 &	0.42\\
hot neptune &	116 &	71 &	29 &	0.41\\
neptune &	51 &	25 &	6 &	0.24\\
mini neptune &	25 &	24 &	7 &	0.29\\
terrestrial &	85 &	67 &	20 &	0.3\\
ice planet &	4 &	0 &	0 &	0\\ \hline \hline
undefined &	516 &	478 &	142 &	0.3\\
unknown &	332 &	304 &	32 &	0.11\\
    
    \end{tabular}
\end{center}
  \caption{\emph{The known exoplanets categories overall and in multiple systems, with the count and fraction of those in/near resonance.}}
\end{table}
\noindent
   It was found that 69\% of the known objects from multiple planetary systems had an inner/outer orbital period ratio between 1 and 3 for the adjacent orbits. Overall 35\% of the objects found in multiplanetary systems had their adjacent orbit ratios within 10\% relative distance of the resonance numbers: 3/2, 5/3 or 2 (these were the most common resonances encountered overall). Even though the analyzed systems are still incompletely known, the low number of occurrences for high orbital period ratios suggest that most planets tend to have stable orbits for orbital period ratios of the adjacent orbits lower than 5. For ratios higher than 3 it is also very likely that some planets (or asteroid belts) will be discovered there in the future to fill in the missing gaps. Another interesting fact is that almost 41\% of the defined planets (i.e. those where the calculations could be made more precisely) are near or in an orbital mean motion resonance.\\
   In order to quickly calculate the resonances or near-resonances of thousands of planet pairs, equations (4) and (5) were used for cases up to the 4th order.  A simplified method was used for orders higher than 4, with the libration width estimated from the equation:
 \begin{equation}    
{\delta n_{max} \over n} = \pm \Bigg({{12 |D_r| e^{q}} \over n}\Bigg)^{1/2},
\end{equation}  
where $D_r$ was calculated empirically to be:
\begin{equation}    
{D_r} =  { {m'} \over {M} } \exp(6-A(r)\cdot r).
\end{equation}
Here $r$ is the resonance number and $A(r)$ another empirical function dependent of it, described by the relation:
\begin{equation}    
{A(r)} = \exp(1.122-1.239 \cdot r) + 1.674.
\end{equation}
 With the planets automatically rearranged by increasing orbital period in each system and the orbital period ratio known, the orbital ratio values were compared to a list of resonance numbers (given by integer number ratios) and matched with the nearest value. Then a test particle was placed onto the orbit of the inside planet. The planar, circular, restricted problem was used, with a body of an external orbit perturbing an inner body of negligible mass, as presented in most detail by Murray and Dermott in ``Solar System Dynamics'' \cite{murray1999solar}. \\
The libration width was calculated as described in section 2. If the planet was inside the libration interval, it was considered in resonance. However, when compared with resonance data obtained through more advanced methods from other articles, it was found that the resonance condition tended to be in general too restrictive. To match the results of this simplified method with those published in other papers while using more accurate techniques, the libration width had to be adjusted. It was found that a value of three times the libration width was generating results that had the best match with data found in other publications. The planet type with which the inside or outside planet was in resonance was neglected in the statistics used here. \\
\noindent
 Because the method assumes that the interior planet has a negligible mass (the libration width can be directly calculated only for internal resonances), it can, of course, generate errors. For this reason, the accuracy of the calculations was compared with over 51 cases of planets near or in resonance that were treated in deep detail and published in various scientific journals. The details of this comparison and the source references used in it are described in the Appendix at the end of the article. It was found that the method produced accurate results in finding out whether a pair of planets is in resonance or near resonance in 86\% of  the situations. However, when the intention was only to estimate if a pair of planets is either in or near resonance, the precision reached 100\%. In consequence, an analysis of both planetary resonances and near resonances in multiplanetary systems was preferred. It was also assumed that the near-resonance cases were at least as important as resonance cases, due to the indications they can give about the evolution of a system from or into a resonant state. And last, but not least, some room was given for possible uncertainties in the accuracy of the existing data that could switch in the future a system from a resonant to a near-resonant state or the other way around. The comparison between the method used in this article and literature is shown in the Appendix at the end of the article.\\
\noindent
Various resonances were investigated up to the 4th order and up to the value 4 for the denominator and calculated using the method described in Section 2.  Resonances of orders higher than 4 were also calculated with the more approximate method described by Eq. (13)-(15) and included in the statistics when encountered. Values higher than 4 for the denominator were not used, in order to avoid the overlapping of libration ranges. \\
The statistical result of the resonance distribution as a function of planet type is shown in Figure 3. The resonant/near resonant brown dwarfs, neptunes and mini-neptunes are the types of planets with the lowest statistics while jupiters and hot jupiters the most common. In spite of the smaller mass, the terrestrial-type planets are almost as numerous as the hot neptunes and more common than the neptune-type planets. While the the objects with larger masses tend to be found more often in systems with single planets, terrestrial-type planets of mini-neptunes are seen more often in multiplanetary systems.\\
 \begin{figure}[t!]
  \centering
  \includegraphics [width=0.99\textwidth]{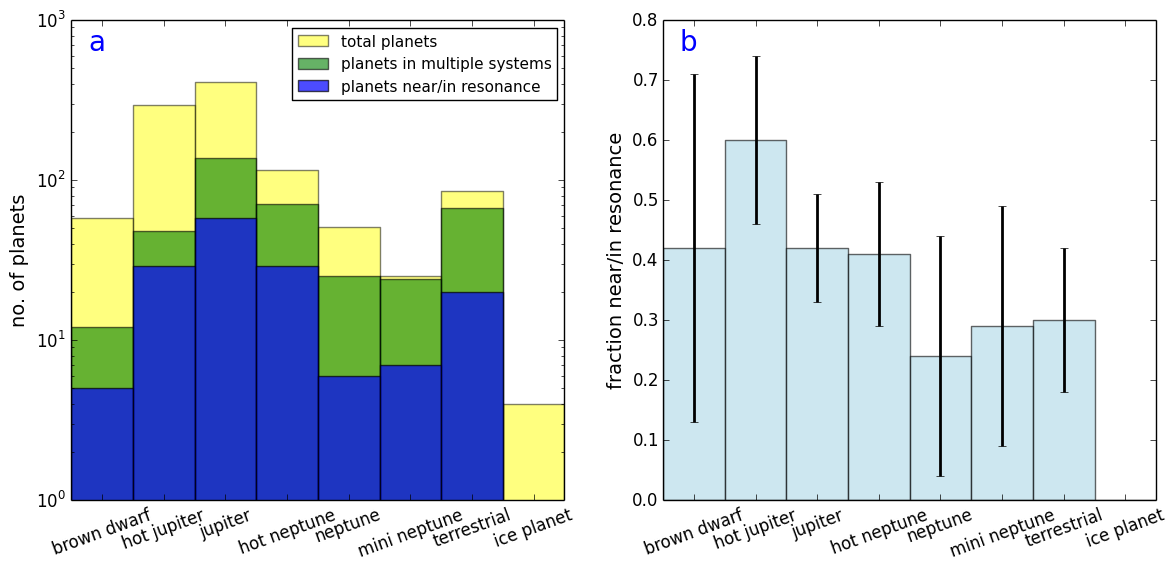}\\
  \caption{\emph{Statistical distribution of exoplanets by type. The yellow bars represent the total no. of analyzed planets, the green bars the planets found in multiple systems and the blue bars the planets near or in resonance (left). The right side of the graph includes their fraction in/near resonance. The statistical error bars for 95\% confidence are displayed as vertical lines.}}\label{Figure-distro.png}
\end{figure} 
\noindent
  The relatively low number of available samples in the observational data generated large error bars for a 95\% confidence in some of the planet categories, however, the results are still statistically significant to compare, for example, jupiter or neptune-type planets with terrestrial planets. The fraction of planets in/near resonance appears to be higher for the categories with larger masses, with planets from categories with masses within comparable ranges having in general a similar fraction in/near resonance, with one exception - the brown dwarfs. However, in this case, the very low statistics still leaves enough room for a significant data change in the light of future discoveries. Due to the fact that the occurrence of the brown dwarfs seems to increase as the mass of the host star decreases, it is also apparent that their formation process differs from that of the lower-mass planets, with these types of planetary objects being actually failed stars, ejected from the stellar embryos taking shape in the initial molecular cloud \cite{{reipurth2001formation}, {bate2002formation}}.\\
\begin{figure}[t!]
  \centering
  \includegraphics [width=0.9\textwidth]{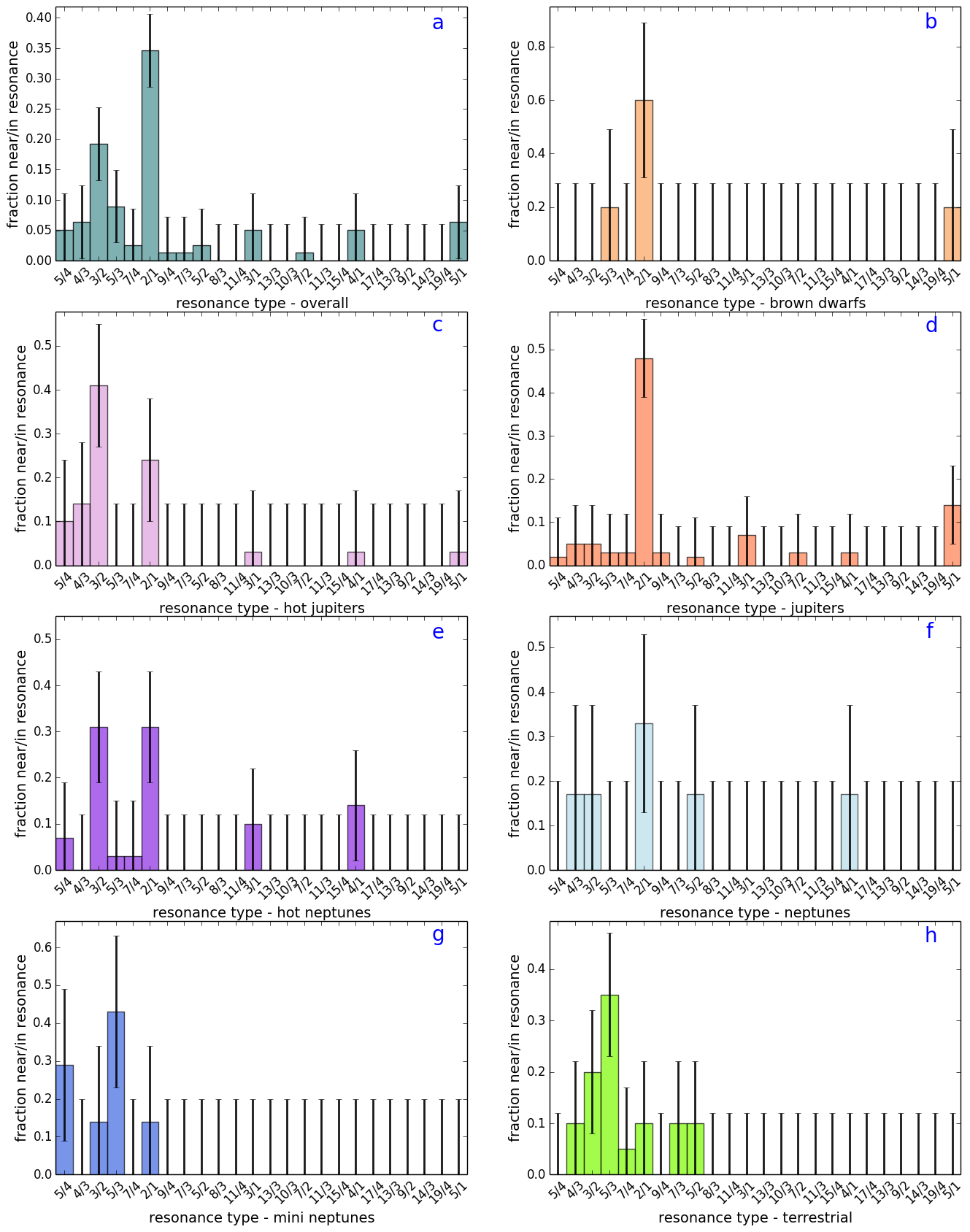}\\
  \caption{\emph{Resonance and near-resonance frequency by planet type. The vertical bars represent the error intervals for 95\% confidence.}}\label{Resonances-by-planet.png}
\end{figure}
A second, normalized statistical distribution of the fraction of planets in or near resonance vs the resonance type is displayed in Figure 4, for the general case in diagram a and then for each planet category in the following graphs. The vertical lines mark again the error bars for 95\% confidence. It is immediately apparent that the resonance or near-resonance distribution is strongly influenced by the planet category. As already specified, the type of the other planet with which the planet of interest was in or near-resonance (internal or external) was neglected.\\
a) It is immediately apparent in Figure 4a that the 2/1 and 3/2 resonances dominate for the general planet distribution, with the first occurring in almost 35\% of the cases while the second appears in over 15\% of the situations. Other significantly statistical values are visible for 5/4, 4/3, 5/3 and for the higher integer values of 3/1, 4/1 and 5/1. From these, only 5/1 is over the background threshold. The 95\% confidence error bars are set here at $\pm$ 6\%\\ 
b) Figure 4b treats the brown dwarf distribution. Due to the very low statistics (only 5 brown dwarfs are in or near resonance), the error bars are very large, at $\pm$ 29\%. However, the 2/1 resonance appears to be statistically significant while one case for each of the values 5/3 and 5/1 was observed.\\
c) The 3/2 resonance for hot jupiters is the dominant value, with over 40\% of the planets near or in this configuration - as shown in Figure 4c. About a quarter of them are encountered in/near the 2/1 configuration. Besides the values of 4/3, resonant/near resonant configurations are also visible at 5/3, 3/1, 4/1 and 5/1, but all below the background threshold of 14\%.\\
d) The jupiter types of planets have the best statistics from all categories, with 68 planets near/in resonance. The 2/1 configuration is dominant here, with over 45\% of the cases observed near this number. Resonance/near resonance values are also encountered for most resonance numbers investigated, but only the 5/1 value is over the background threshold limit of 9\%. What is remarkable is the very low occurrence of the 3/2 values, most common for the hot jupiters. The graph is displayed in Figure 4d.\\
e) The hot neptunes have the 3/2 and 2/1 resonances almost equally represented - slightly higher than 30\%, with other significant values at 4/1, just above the  background threshold of 12\%, and 3/1 just below it. Much lower than the threshold value, resonances/near resonances at 5/4, 5/3 and 7/4 are also present (Figure 4e). \\
f) The neptune-type planets (Figure 4f) exhibit one significant resonance/near resonance fraction at almost 35\%, for the value of 2/1, with other values at 4/3, 3/2, 5/2, 4/1 and 5/1 slightly below the background threshold of 20\%.\\
g) The mini-neptunes in evidence have 5/3 as the most important resonance number in the distribution, at almost 45\%, and 5/4 at almost 30\%, both above the 20\% threshold limit. Other still statistically significant values are seen at 3/2 and 2/1, at around 15\% (Figure 4g).\\
h) The terrestrial planets (Figure 4h) appear to have a somewhat continuous resonance distribution, just below the threshold of 12\% for most cases, with all numbers except 9/4  encountered between 4/3 and 5/2. They present a significant increase, at almost 35\%, for the value of 5/3 and a drop to about 5\% at 7/4. More data will be needed here, too, for a confident statistical characterization of their resonance/near resonance profile, but it is interesting to remark that the 3/2 and 2/1 values that dominate the distribution of most larger- mass planet categories seem to be notably less common in this case.\\
As a short conclusion, it can be said that the resonance/near resonance numbers of 2/1 and 3/2 appear to be more common for the planets with larger masses while the 5/3 resonance seems to be dominant for terrestrial planets and mini neptunes. For giant planets, the 2/1 resonances are more common at larger distances from the host star while the 3/2 resonance is better represented at close distances from it. Resonances for values higher than 5/2 are encountered only for planets with masses larger than ($M_E$). \\
\begin{figure}[htb!]
  \centering
  \includegraphics [width=0.9\textwidth]{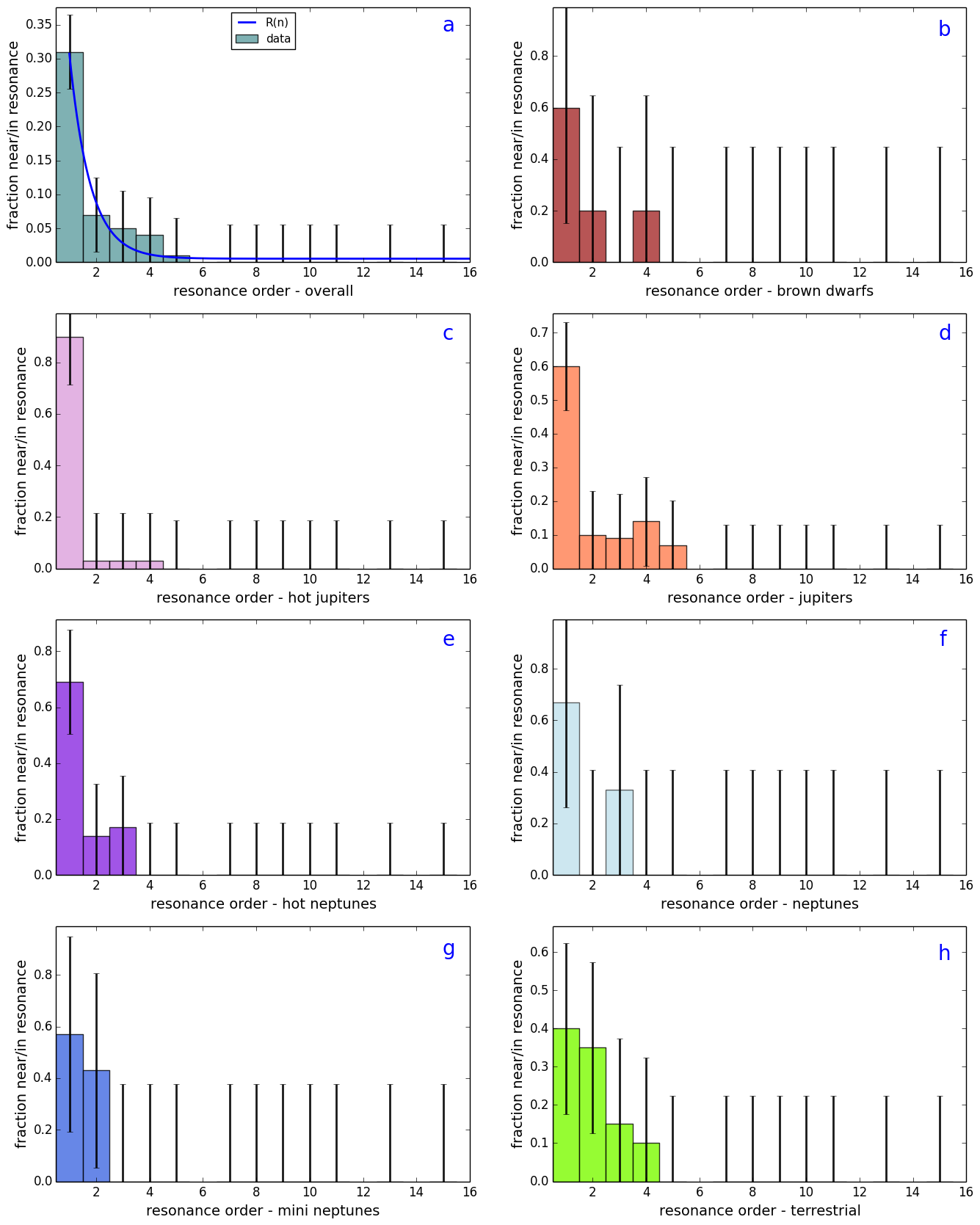}\\
  \caption{\emph{The fraction of planets in/near resonance by resonance order, displayed for each planet category. The vertical bars represent the error intervals for 95\% confidence.}}\label{Resonances-by-order.png}
\end{figure}
In a further graph (Figure 5), the the fraction of resonances/near resonances occurrences were also analyzed as a function of their order, with the statistical distribution error for 95\% confidence displayed again as vertical bars. Due to the rapid decrease of the libration width with the resonance order, a quick decrease in the planetary distribution was also expected for all categories. Most observational data confirmed that the lower order resonances were in general much more common than the higher orders, with the resonances of the first order dominating the overall distribution. \\
a) Figure 5a shows that the overall planetary distribution follows an exponential-type law of the resonance/near resonance fraction as a function of the resonance order, with first order dominating at almost 30\%. The correlation of the fraction in/near resonance vs the resonance order $n$ can be described by an empirical exponential function:
 \begin{equation}       	
R(n)=1.268 \cdot \exp(-1.437 n) + 0.006,
\end{equation}
b) The distribution for brown dwarfs suggests another decrease in the resonance order, but not as steep as for the overall planets, as shown in Figure 5b. The first order resonances have a frequency of occurrences of around 60\%. However, the low statistics makes this set of data less reliable.\\
c) The hot jupiters also appear follow an exponential decrease, with a very steep drop past the first resonance order (Figure 5c). The first order resonances account to about 90\% of occurrences. \\
d) The jupiter-type planets follow generally an exponential decrease in their distribution, similar to the overall function displayed in Figure 5a as well, however, no resonances/near resonances of the 3rd order have been observed (Figure 5d). First order resonances/near resonances make up almost 60\% of occurrences.\\
e) The hot neptunes also may follow an apparent exponential decay, with almost 70\% of resonance or near-resonance occurrences belonging to the first order (Figure 5e).\\
f) The neptune-type planets also follow an exponential law for the resonance/near resonance distribution with the first order approaching 60\%. No second order resonances/near resonances have been observed, however the low data statistics leaves room open for any possibility (Figure 5f). The low statistics also lives room for many possible changes as the number of the planet in this category is expected to increase in the future.\\
g) The statistics of the mini-neptunes are also low, with only resonances of the first and second type encountered, their distribution also possibly following an exponential decrease with a lower decay rate than for the overall resonance distribution (Figure 5g). \\
h) The resonance/near resonance fraction for the terrestrial planets has a less steep decrease, with the first order at around 42\%. The function fitting it might be an exponential or a Gaussian, the data being insufficient for good accuracy.  (Figure 5h). \\
What is worth remarking is that only the Jupiter-type planets have a small fraction in a 5th order resonance/near resonance (they also are best represented statistically), all the other categories not going beyond the 4th order with the available observational data. Orders higher than 5th are not encountered for any planet category. The decrease of the resonance percentage by order appears to be sharper for giant planets that are orbiting close to the host star than for those orbiting further away. The mini neptunes and the terrestrial planets appear to have a slower decrease of the resonance fraction as a function of the resonance order. More data, with better statistics, will be needed for a clear conclusion. \\
\begin{figure}[htb!]
  \centering
  \includegraphics [width=0.9\textwidth]{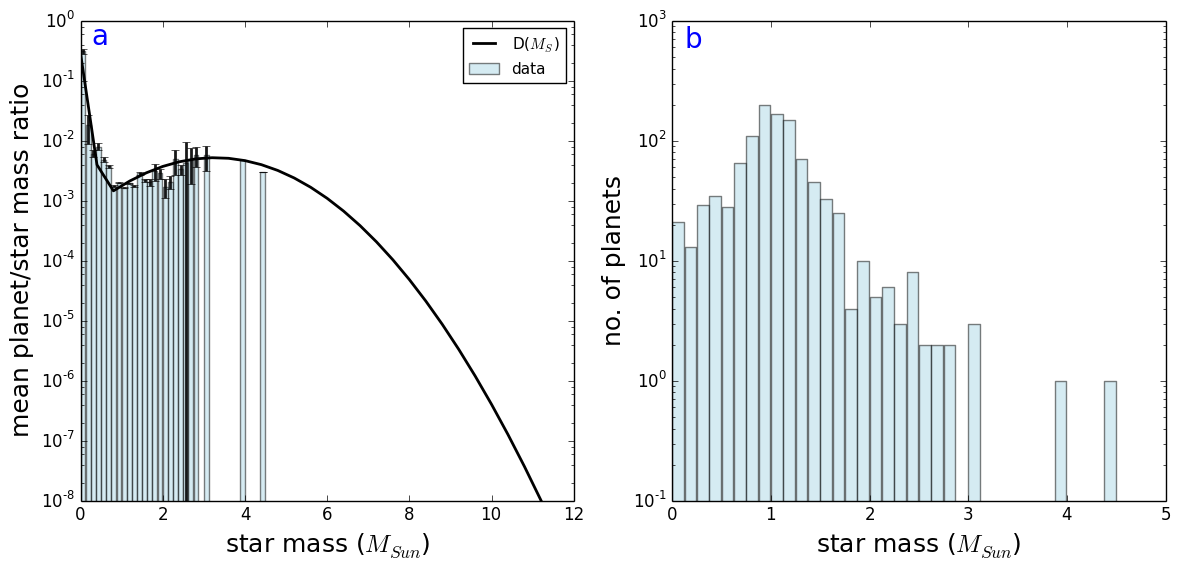}\\
  \caption{\emph{Distribution function for the planet mass/star mass ratio of the defined planets (left). The vertical lines represent the statistical error bars for 95\% confidence. No error bars are shown for 4 and 4.5 $M_{Sun}$, because of the existence of a single available data sample for each of them. The continuous line shows the fitting derived function of these data. The right side of the figure shows the number of samples (planets) for the corresponding star mass. }}\label{Distro.png}
\end{figure}
\noindent
With an important number of planets unknown and undefined (i.e. the planet mass is unknown or the host star mass and/or the planet mass is unknown), an attempt was made to estimate the objects from these categories in or near-resonance, too. First, a distribution for the planet mass/star mass ratio as a function of the star mass expressed in solar masses was derived from the available data (Figure 6). The distribution was then empirically fit with a combination of exponential and Gaussian functions described by the expression:
 \begin{equation}       	
D(M_S)=0.267 \cdot \exp(-14.899 M_S) + 0.00263+0.005\cdot {{\exp(M_S-3.261)^2}\over {6.864}}.
\end{equation}
Obviously, the almost inexistent data for star masses higher than 3 $M_{Sun}$ makes the Gaussian approximation very uncertain, but for the time being it is the the best fit that can be obtained. The discontinuity around 0.3 star masses appears to be a threshold factor from where the planet formation process changes. For very low-mass stars, the planet/star mass ratio appears to increase exponentially, apparently the low-mass proto-star nebula is quickly absorbed by high-mass planets (jupiter-type of brown dwarfs ejected from the proto-core), while for higher-mass stars it is possible that the radiation pushes away most of the nebula material, decreasing the final mass of the formed planets. \\
\begin{figure}[htb!]
  \centering
  \includegraphics [width=0.9\textwidth]{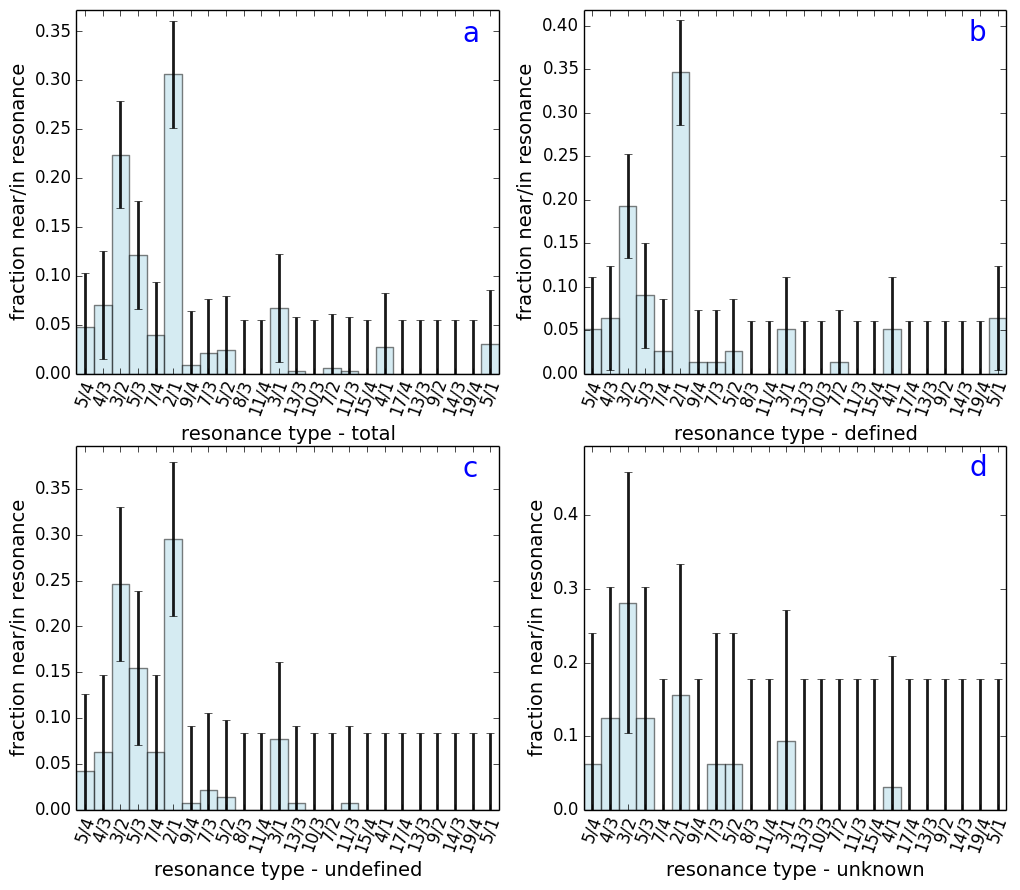}\\
  \caption{\emph{The fraction of planets in/near resonance for the general planet distribution and then, separately, for defined, undefined and unknown planets. The error bars for 95\% confidence are marked as vertical lines.}}\label{Resonances-by-defined.png}
\end{figure}
\noindent
The distribution from Figure 6 was used to estimate the average ratio of the planet mass over the host star mass for the undefined planets and then used in Equations (3)-(5). For the unknown types of planets, the average ratio was taken as $10^{-4}$, close to the average value of the planet mass ratio for the cases with known masses, and also applied to Equations (3)-(5). The accuracy of the method was tested and compared for the undefined and defined planets. No differences were observed if the distribution function was used instead of the real data. For the unknown planets, the overall difference from the defined planets was within 1\%, however there were important fluctuations on some the planet categories, with a tendency of the resonance/near resonance fraction to increase up to 30\% for the lower-mass mini-neptunes and terrestrial planets. \\
\begin{figure}[htb!]
  \centering
  \includegraphics [width=0.9\textwidth]{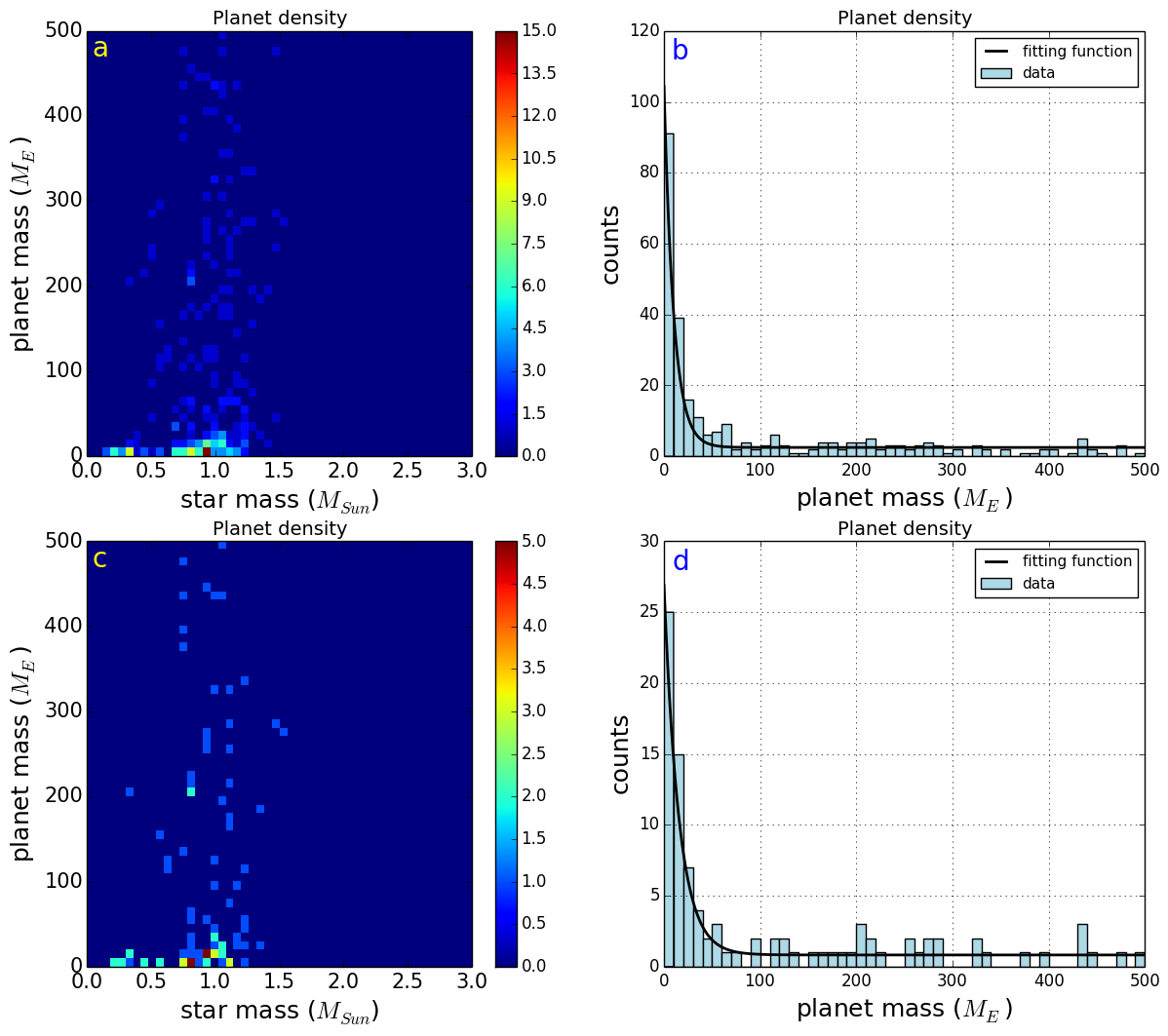}\\
  \caption{\emph{Planet mass vs star mass for general cases and near-resonance objects up to 500 ($M_E$). The mass distributions in figures b and d fit well with exponential-type functions of the form $a \cdot exp(-bm)+c$.}}\label{planet-star-mass.png}
\end{figure}
\noindent
All the data from multiplanetary systems objects was added and compiled for an overall distribution and also plotted in separate graphs for the defined, undefined and unknown planets, with the error bars for 95\% confidence marked ( Figure 7). The overall resonance/near resonance distribution (Figure 7a) shows again an affinity for the 2/1 and 3/2 resonances at almost 31\%, respectively almost 23\%, with some other notable occurrences at 5/3, 4/3 and 3/1. Figure 7b is identical with Figure 4a and represents the defined planets, where the resonances are most accurately determined. Here, resonances/near resonances at 2/1 (35\%) and 3/2 (20\%)are dominating. The same types of resonances/near resonances are the most common for the undefined planets, but the 3/2 resonance/near resonance is observed in 25\% of the cases while about 30\% of the objects are in the vicinity of the 2/1 resonance/near resonance. Resonances/near resonances higher than 11/3 are not observed, but this fact may be due to the approximations coming from the distribution function. The ``unknown'' class has a different profile, with the 3/2 ratio dominating at almost 28\%, followed by 2/1 for much lower values at around 16\%. However, the approximations used here might change significantly the distribution from the real data. \\
While for the unknown type of planets the 2/1 resonance appears to be the most common and the only one above the confidence threshold, approaching 30\% (but the data is too arbitrary), for the undefined and defined planets it's obvious that the 2/1 resonance is the most important, followed by the 3/2 value. The data for the defined planets remains the set that can be treated by far with the highest level of confidence.

\section{Planet mass vs resonance/near resonance probability}

The exoplanet distribution as a function of star and planet mass was also studied for both the general case and the case of planets in/near resonance (Figure 8). it is interesting to observe that the planet count in the histograms appears to follow an exponentially decreasing function for both cases in the range of 0-200 ($M_E$). The exponential function fitting the general data is described by the equation:\\
 \begin{equation}       	
e_G(m)=92.49 \exp(-0.091 m) + 2.43,
\end{equation}
while the function fitting resonance/near resonance data has the form:\\
 \begin{equation}       	
e_R(m)=23.43 \exp(-0.054 m) + 1.04,
\end{equation}
From Equations (14) and (15) one can estimate the resonance fraction as a function of the planet mass:\\
 \begin{equation}       	
F(m)={{e_R(m)}\over{e_G(m)}}.
\end{equation}
The relatively low statistics from the available data makes Equation (16) more of a tentative attempt to describe the resonance fraction distribution, which should be regarded with caution. With the count for the lower mass planets much higher, the resonance/near resonance distribution was plotted in Figure 9 for both lower mass planets and the full range of the planetary masses present in the available data.  
\begin{figure}[t!]
  \centering
  \includegraphics [width=0.9\textwidth]{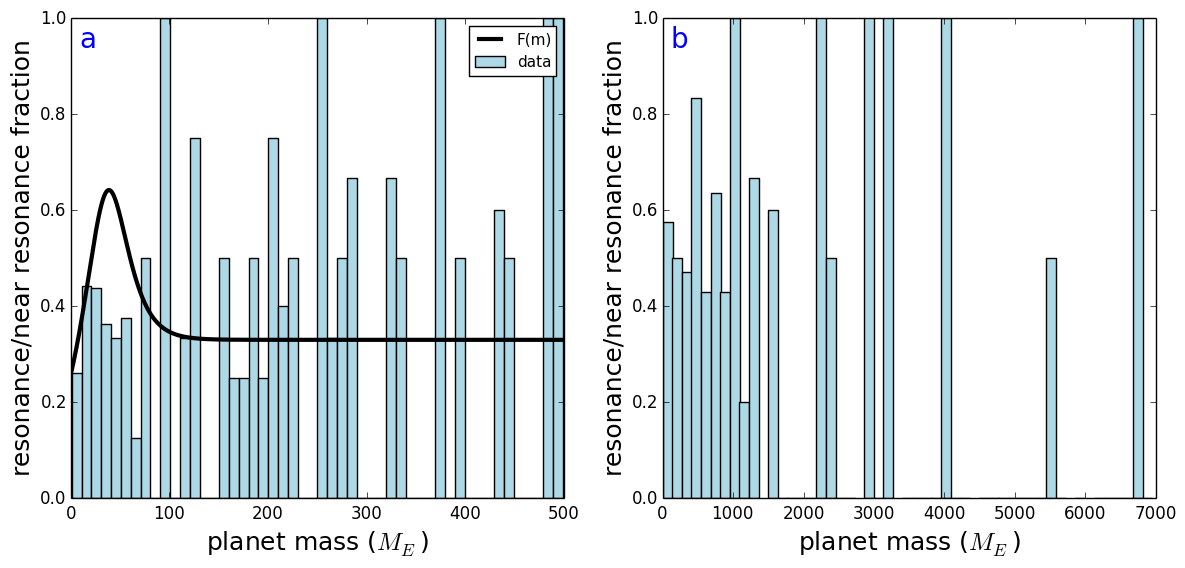}\\
  \caption{\emph{The fraction of planets near/in resonance as a function of their mass.The plot on the left side shows the distribution for lower masses while the right side encloses the full range of planetary masses observed near/in resonance. The line representing F(m) should be considered only up to about 200 $M_E$.}}\label{mass-res-combined.png}
\end{figure}
While no obvious correlation between planet mass and resonance/near resonance fraction could be derived directly from the plotted histograms, the distribution $F(m)$, also displayed in Figure 9a, suggests the possibility of a rise and fall in the planet resonance/near resonance fraction, with a peak at around 50 $M_E$, followed by a decrease to an almost uniform value of 0.45 at around 200 $M_E$. For higher mass values, the histogram from Figure 9b suggests a fraction value of around 0.55-0.6 up to 1500 $M_E$. Mass values over 1500 $M_E$ have even lower statistics and the resonance/near resonance fraction distribution becomes totally undetermined. This result might imply that while lower mass planets do not have enough gravitational field to induce resonances or near-resonances, the higher mass planets might absorb too much matter during the accretion stage and may become part of a system with fewer major planets, decreasing the probability of a resonant or near-resonant orbital configuration. Planets with much higher mass, approaching the brown dwarf lower-mass limit, are most likely taking shape through a different process and this might affect their affinity for orbits in/near mean motion resonance. More data will be necessary before Equation (20) can be confirmed, infirmed, or adjusted. The only correlation that appears to exist so far is a correspondence between the planet mass/star mass ratio and star mass, as shown in the previous section, in Figure 6.

\section*{Conclusions}

A statistical analysis of the orbital parameters for the exoplanets from multiplanetary systems can be potentially very useful to determine the general mechanisms of planet formation and to improve the existing theoretical models and numerical simulations. With a method of rapid calculation of the snow line for every host star, it appears that the planets' density plot as a function of semimajor axis/snow line and planet mass tends to be organized in three main categories that have higher density concentrations. These areas are apparently related to different mechanisms of planet formation, with the regions in between subjected to a combination of them. Based on this data, a slightly adjusted classification of the known exoplanet categories is suggested.\\
Performing a simple analysis, the resonance or near-resonance states present in all the multiplanetary systems known to date can be found numerically using a computer analysis tool. The first results, presented in this paper, suggest different resonance or near-resonance distributions for different planet categories. The resonance/near resonance numbers of 2/1 and 3/2 appear to be dominant for the planets with larger masses while the 5/3 resonance seems to be the most common for terrestrial planets and mini neptunes. For giant planets, the 2/1 resonances are dominating at larger distances from the host star while the 3/2 resonance is more common at close distances from it. Resonances for values higher than 5/2 are encountered only for planets with masses larger than 10 ($M_E$). When the resonance/near resonance distribution is studied as a function of the resonance order, the decrease of the resonance number percentage appears to be sharper for giant planets that are orbiting close to the host star than for those orbiting further away. The mini neptunes and the terrestrial planets display a slower decrease of the resonance fraction as a function of the resonance order.\\
A more approximate attempt to make estimates for the exoplanet distribution by resonance number was made for undefined and unknown planets, using a distribution function for the planet/star mass ratio from the defined planets data for the first category and an approximate mass ratio of $10^{-4}$ for the second. The results in the first case appear similar to those from the defined category while in the second case they look visibly different. One can conclude that the approximate method that uses the mass ratio distribution function can bring satisfactory results while the fixed number approximation should be treated with caution. \\
An attempt to derive a relationship between the planet mass and its resonance/near resonance fraction was also made by approximating the general and resonant/near resonant distributions with two exponential functions. The resulting equation suggests a resonance peak of about 0.75 at around 50 $M_E$, followed by a decrease to an almost uniform value of 0.45 at around 200 $M_E$ while the distribution for higher masses suggests higher values of 0.55-0.6 up to 1500 $M_E$. \\
As more exoplanets will continue to be discovered in the future, the statistics for these types of calculations are expected to improve significantly. More specialized studies based on certain star or planet types or on certain physical properties of the studied multiplanetary systems might bring even more interesting results in this field and they are scheduled to be treated in future publications. 

\subsubsection*{Aknowledgnements}
This research has made use of the Extrasolar Planet Encyclopedia Archive, which is operated by l'Observatoire de Paris and of the NASA Exoplanet Archive, which is operated by the California Institute of Technology, under contract with the National Aeronautics and Space Administration under the Exoplanet Exploration Program.

\section*{Appendix}

The results of the simplified resonance/near resonance detection method used in this paper was compared with those obtained using more sophisticated tools and presented in other publications. They were found to be sufficiently accurate for the purpose of this article. Table 2 presents the comparison between the two sets of data. The resonance numbers displayed are the orbital period ratio of the outer/inner adjacent planets. The analyzed cases were those of resonance (R), near-resonance (NR), probable resonance (PR), and probable near undefined resonance (PNUR). The resonances are considered related to the inner (in) and outer (out) planet.

 \LTcapwidth=\textwidth 

  \begin{longtable}{ p{2.8cm} p{2.4cm} p{1.9cm} p{2.9cm} p{3.3cm} }

\hline
\textbf{planet name \newline} &\textbf{planet type} & \textbf{resonance\newline type} & \textbf{simplified \newline method} & \textbf{literature} \\ \hline

\endhead
\hline \multicolumn{5}{r}{\textit{Continued on next page}} \\
\endfoot 
\endlastfoot

24 Sex b &	jupiter &	na &	R out &	R out \cite{Johnson:2010zw} \\
24 Sex c &	jupiter &	2/1 &	R inn &	R inn \cite{Johnson:2010zw} \\
55 Cnc b &	hot jupiter &	$>$5 &	R out &	PR out \cite{nelson2014remastering} \\
55 Cnc c &	hot neptune &	3/1 &	R inn &	PR inn \cite{nelson2014remastering} \\
BD20 2457 b &	brown dwarf &	na &	R out &	PR out \cite{horner2014dynamical} \\
BD20 2457 c &	jupiter &	5/3 &	R inn &	PR inn \cite{horner2014dynamical} \\
Gliese 876 c &	jupiter &	$>$5 &	R out &	R out \cite{rivera2010lick} \\
Gliese 876 b &	jupiter &	2/1 &	R inn/NR out &	R inn/out \cite{rivera2010lick} \\
Gliese 876 e &	neptune &	2/1 &	NR inn &	R inn \cite{rivera2010lick} \\
HD 10180 d &	hot neptune &	11/4 &	NR out & NR out \cite{lovis2011harps} \\
HD 10180 e &	hot neptune &	3/1 &	NR inn &	NR inn \cite{lovis2011harps} \\
HD 128311 b &	jupiter &	na &	R out &	R out \cite{rein2015reanalysis} \\
HD 128311 c &	jupiter &	2/1 &	R inn &	R inn \cite{rein2015reanalysis} \\
HD 155358 b &	jupiter &	na &	R out &	R out \cite{robertson2012mcdonald} \\
HD 155358 c &	jupiter &	2/1 &	R inn &	R inn \cite{robertson2012mcdonald} \\
HD 200964 b &	jupiter &	na &	R out &	R out \cite{wittenmyer2012resonances} \\
HD 200964 c &	jupiter &	4/3 &	R inn &	R inn \cite{wittenmyer2012resonances} \\
HD 202206 b &	brown dwarf &	na &	R out &	R out \cite{couetdic2010dynamical} \\
HD 202206 c &	jupiter &	5/1 &	R inn &	R inn \cite{couetdic2010dynamical} \\
HD 204313 b &	jupiter &	$>$5 &	R out &	R out  \cite{robertson2012second}\\
HD 204313 d &	jupiter &	3/2 &	R inn &	R inn  \cite{robertson2012second}\\
HD 37124 c &	jupiter &	5/1 &	R out &	PR out \cite{wright2011california}\\
HD 37124 d &	jupiter &	2/1 &	R inn &	PR inn \cite{wright2011california} \\
HD 40307 d &	terrestrial &	2/1 &	NR out &	NR out \cite{tuomi2013habitable} \\
HD 40307 e &	terrestrial &	5/3 &	NR inn/R out &	NR inn/out - \cite{tuomi2013habitable} \\
HD 40307 f &	terrestrial &	3/2 &	R inn &	NR inn \cite{tuomi2013habitable} \\
HD 45364 b &	neptune &	na &	R out &	R out \cite{correia2009harps} \\
HD 45364 c &	jupiter &	3/2 &	R inn &	R inn \cite{correia2009harps} \\
HD 60532 b &	jupiter &	na &	R out &	R out \cite{laskar2009hd} \\
HD 60532 c &	jupiter &	3/1 &	R inn &	R inn \cite{laskar2009hd} \\
HD 73526 b &	jupiter &	na &	R out &	R out \cite{wittenmyer2014detailed} \\
HD 73526 c &	jupiter &	2/1 &	R inn &	R inn \cite{wittenmyer2014detailed} \\
HD 82943 c &	jupiter &	na &	R out &	R out \cite{tan2013characterizing} \\
HD 82943 b &	jupiter &	2/1 &	R inn &	R inn \cite{tan2013characterizing} \\
HR 8799 c &	jupiter &	2/1 &	R inn &	R out \cite{fabrycky2010stability} \\
HR 8799 b &	jupiter &	2/1 &	R inn &	R inn \cite{fabrycky2010stability} \\
KOI-82 c &	terrestrial &	3/2 &	NR out &	NR out \cite{wu2013density} \\
KOI-82 b &	terrestrial &	3/2 &	NR inn &	NR inn \cite{wu2013density} \\
Kapteyn's b &	terrestrial &	na &	R out &	NR out \cite{anglada2014two} \\
Kapteyn's c &	terrestrial &	5/2 &	R inn &	NR inn \cite{anglada2014two} \\
Kepler-9 b &	hot neptune &	$>$5 &	NR out &	NR out \cite{holman2010kepler} \\
Kepler-9 c &	hot neptune &	2/1 &	NR inn &	NR inn \cite{holman2010kepler} \\
NN Ser (ab) d &	jupiter &	na &	R out &	R out \cite{beuermann2010two} \\
NN Ser (ab) c &	jupiter &	2/1 &	R inn &	R inn \cite{beuermann2010two} \\
PSR 1257 12 c &	unknown &	8/3 &	PNUR out &	NR out \cite{petrovich2013planets} \\
PSR 1257 12 d &	unknown &	3/2 &	PNUR inn &	NR inn \cite{petrovich2013planets} \\
mu Ara d &	jupiter &	$>$5 &	R out &	R out \cite{mushailov2014analytical} \\
mu Ara b &	jupiter &	2/1 &	R inn &	R inn \cite{mushailov2014analytical} \\
ups And c &	jupiter &	$>$5 &	R out &	R out \cite{mushailov2014analytical} \\
ups And d &	jupiter &	5/1 &	R inn &	R inn \cite{mushailov2014analytical} \\

\\

\caption{\emph{Resonant/near resonant behaviour of exoplanets calculated using the simplified method of external resonances compared with methods used in other publications}}

\end{longtable}    
\vspace{10pt}


\end{document}